\documentclass[%
 reprint,
 amsmath,amssymb,
 aps,
]{revtex4-1}

\usepackage{graphicx}
\usepackage{algorithm}
\usepackage{amsmath}
\usepackage{amsfonts}
\usepackage{array}
\usepackage{dsfont}
\usepackage{multirow}
\usepackage{algpseudocode}
\usepackage{hyperref}   
\graphicspath{{figures/}}
\usepackage{dcolumn}
\usepackage{bm}

\begin{document}

\title{Efficient and fast algorithms to generate holograms for optical tweezers}

\author{Naomi Holland}
\author{Dustin Stuart}
\author{Oliver Barter}
\author{Axel Kuhn}
 \email{axel.kuhn@physics.ox.ac.uk}
\affiliation{
Clarendon Laboratory, University of Oxford, Parks Road, Oxford, OX13PU}

\date{\today}

\begin{abstract}
We discuss and compare three algorithms for generating holograms: simple rounding, Floyd-Steinberg error diffusion dithering, and mixed region amplitude freedom (MRAF). The methods are optimised for producing large arrays of tightly focused optical tweezers for trapping particles. The algorithms are compared in terms of their speed, efficiency, and accuracy, for periodic arrangements of traps; an arrangement of particular interest in the field of quantum computing. We simulate the image formation using each of a binary amplitude modulating digital mirror device (DMD) and a phase modulating spatial light modulator (PSLM) as the display element. While a DMD allows for fast frame rates, the slower PSLM is more efficient and provides higher accuracy with a quasi-continuous variation of phase. We discuss the relative merits of each algorithm for use with both a DMD and a PSLM, allowing one to choose the ideal approach depending on the circumstances.

\end{abstract}

\pacs{Valid PACS appear here}
\maketitle

\section{Introduction}

Since their invention by Arthur Ashkin \cite{Ashkin:86}, optical tweezers have had an enormous impact in diverse fields from biology to quantum physics (see \cite{Ashkin2000} for a review). The underlying mechanism is the optical gradient force, which acts on polarisable particles such as living cells \cite{enlighten95082}, nanoscale tools \cite{enlighten62628} or single atoms \cite{PhysRevX.4.021034}, causing them to be trapped at the point of highest intensity of a tightly focused light beam. The potential energy of such a particle is proportional to the intensity of the trapping light at the position of the particle. 

Much effort has been devoted to designing dynamic potential landscapes for trapping and moving large numbers of particles. Regular arrays of thousands of potential wells have been created using optical lattices \cite{NatPhys.3.556, Nature.462.74}, microlens arrays \cite{PhysRevLett.105.170502} and diffractive optical elements \cite{PhysRevA.88.013420}. However, these methods are limited in that the trapping sites can only be moved in unison, not individually. A second approach is to use an acousto-optic deflector (AOD) to generate a steerable trapping beam which can be used to `paint' time-averaged potentials \cite{NewJPhys.11.043030, 0953-4075-41-21-211001, Trypogeorgos:13}. In one demonstration \cite{Opt.Lett.39.2012} the authors generate 32 movable trapping beams using frequency shift key modulation. However, the time averaging is only valid when the oscillation frequency of the trapped particles is much lower than the rate of frequency shifting, which is ultimately limited by the rise time of the AOD.

A third, more flexible approach is to use a spatial light modulator (SLM) to create the desired potential landscape by displaying a hologram which is converted into the desired intensity landscape after propagation through the optical system \cite{NewJPhys.14.073051, RevSciInst.83.11, dustinPaper}. This holographic technique concentrates a large fraction of the optical power in the active trapping sites and allows for three-dimensional positioning, but requires one to calculate the hologram. There are two broad categories of SLMs: phase modulators (PSLMs) such as ferroelectric modulators \cite{JModOpt.51.2235} and liquid crystal displays \cite{Hermerschmidt2008}, and amplitude modulators such as digital mirror devices (DMDs) \cite{Mirhosseini:13}. Holographic optical tweezers typically use a PSLM, with an iterative phase retrieval algorithm such as the Gerchberg-Saxton algorithm \cite{Optik.35.237}, mixed-region amplitude freedom (MRAF) \cite{Pasienski:08}, offset MRAF \cite{mraf2}, or conjugate gradient minimisation \cite{tHarteCGM} to calculate the hologram.

Here we demonstrate the use of both a PSLM, which is quasi-continuous with $m>200$ phase levels between $0$ and $2\pi$, and a binary amplitude modulating DMD, in a holographic optical tweezers arrangement. We consider three algorithms for hologram generation: one from the class of iterative algorithms mentioned above, and two that are considerably faster. We point out briefly here one notable omission from the algorithms we consider, namely conjugate gradient minimisation. This algorithm has recently gained attention for use with PSLMs \cite{tHarteCGM,otherCGM}. However, its main selling point is that it offers simultaneous control of the amplitude and phase of the potential landscape. Since we are only interested in the accuracy of the algorithms in terms of amplitude, conjugate gradient minimisation does not fit well into this discussion and would not be done justice by the comparison.

We show how to use a DMD or PSLM to holographically generate large arrays of individually movable trapping sites. We begin with a brief overview of the principles of holographic imaging. Next, we describe several different algorithms for rapidly calculating artificial holograms for a phase or amplitude modulating device, and show how to apply these to physical modulators that either permit binary amplitude modulation \textit{or} quasi-continuous phase modulation. Finally, we compare the different algorithms on the basis of speed of computation, efficiency of use of laser power, and accuracy of the resulting trapping potentials.

\section{Principles of Holographic Imaging}

\begin{figure*}[tb]
\includegraphics[width=0.9\textwidth]{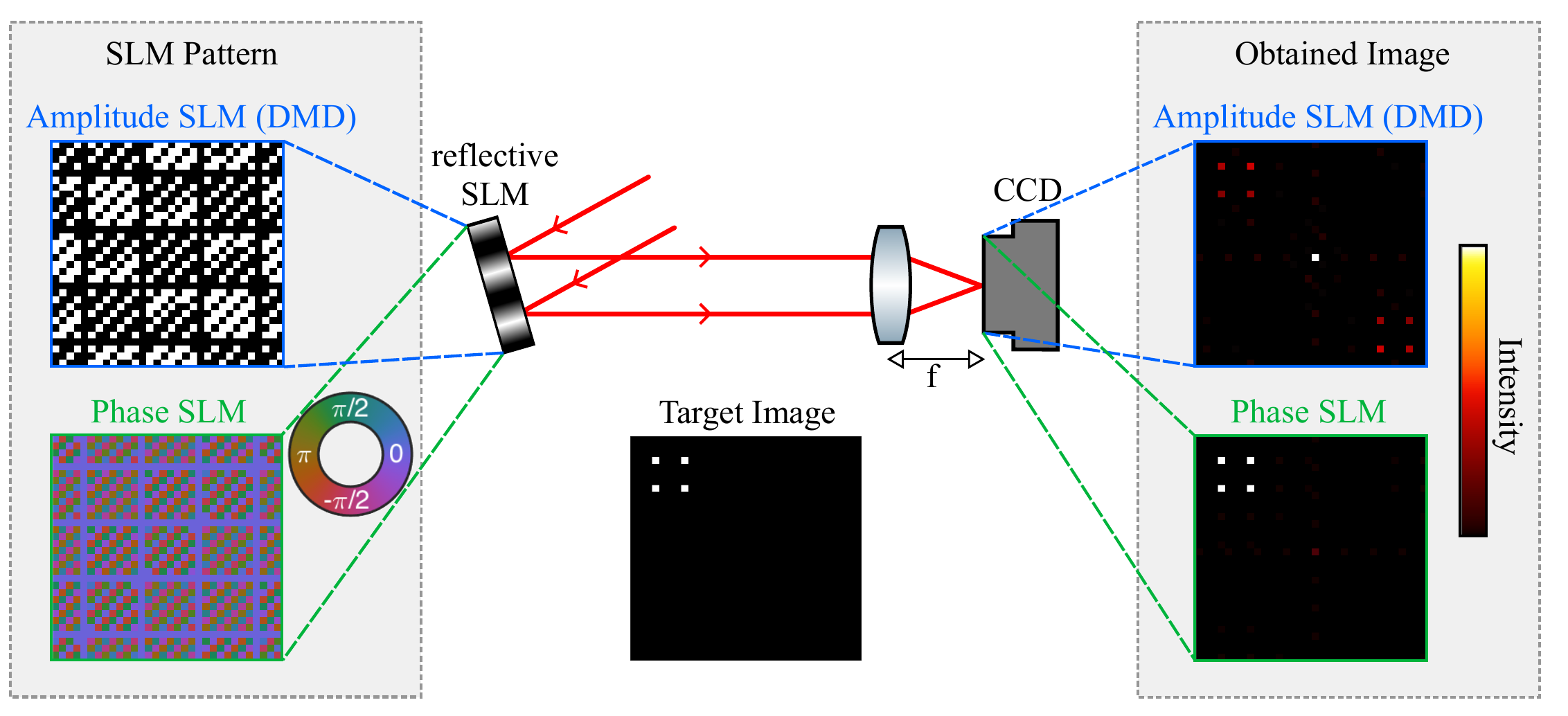}
\caption[justification=justified]{The basic experimental setup required for testing the holographic optical tweezers. A collimated beam of laser light is incident on the SLM. The light after the SLM is focused through a lens to form traps in the focal plane, where a CCD camera is placed to image the traps. In the case of a DMD, only the light in the `1' state would be sent through the lens; light in the `0' state would be reflected out of the setup. [Left]: Sample hologram patterns corresponding to a target image consisting of a $2 \times 2$ grid of traps, displayed on a $32 \times 32$ pixel device for simplicity. [Right]: The actual image planes generated by each of the holograms. In the case of the DMD, we see light in the $-1$,$0$ and $+1$ diffraction orders, while for the PSLM the power all goes to the $+1$ order.}
\label{fig:setup}
\end{figure*} 

The general idea in holographic imaging is to artificially produce an optical field $H(x,y)$, the hologram, which after propagation through the optical system results in a desired optical field $F(x^\prime,y^\prime)$, the image. The problem can be divided into two parts \cite{goodman1996introduction}: the computational problem of how to calculate the required optical field; and the representational problem of how to display the complex-valued field using a physical light modulator. The latter modulates either amplitude \textit{or} phase, which is far from ideal and normally results in artifacts when forming images \cite{artifacts}. The mitigation of these is what necessitates the elaborate algorithms discussed here.

\subsection{The Hologram of a Single Trap} 

The standard experimental setup used for holographic optical tweezers is shown in figure \ref{fig:setup}. The image $F(x^\prime,y^\prime)$ may be calculated from the hologram $H(x,y)$ using the Fresnel diffraction integral \cite{goodman1996introduction}. If we form the image in the focal plane of the lens, this simplifies to a Fourier transform. Hence, given a desired image plane situated in the focal plane of the lens, the required hologram is given by the inverse Fourier transform of this image plane.

We now consider the hologram field of a single diffraction-limited trap

\begin{equation}
H(x,y) = A \exp\Big{(}i\frac{2\pi}{f \lambda}\big{(}x_0^\prime x + y_0^\prime y + \frac{z_0^\prime}{2f}(x^2 + y^2)\big{)}\Big{)},
\label{eq:hologram}
\end{equation}

\noindent where $A$ encodes the trap amplitude and phase, $x_0^\prime, y_0^\prime, z_0^\prime$ are its coordinates relative to the focal point of the lens of focal length $f$, $\lambda$ is the wavelength of the light, and $x$ and $y$ are the coordinates in the plane of the modulator. The two terms linear in $x$ and $y$ account for the lateral position of the trap, while the quadratic $(x^2 + y^2)$ term introduces a small defocus which allows the trap to be moved in and out of the focal plane by a distance $z_0^\prime$. Neglecting the quadratic term, the image plane is a Fourier transform of $H(x,y)$, and the resulting $F(x^\prime,y^\prime)$ is a delta function spot at the location $(x_0^\prime,y_0^\prime)$. In reality, the spot is an Airy pattern whose width is determined by the limiting aperture of the setup. Furthermore, since only the intensity of the image $|F(x^\prime,y^\prime)|^2$ is relevant for trapping, we are free to choose any phase $\arg(A)$ for the trap.

For multiple traps, we extend the above as follows. We want to set the magnitudes of all traps to be equal, such that the total power is distributed evenly. Additionally, we wish to set the phase of each trap to a random value between $0$ and $2\pi$. This is done to prevent the amplitude maxima of all of the individual trap holograms from constructively interfering, which we will see would exacerbate the problems caused by the limitations of a physical light modulator. Additionally, it helps to avoid systematic near-field coherence effects close to the image or focal plane, such as the period doubling in the Talbot-Lau effect \cite{talbot}.

\subsection{Representing a Complex-Valued Hologram}
\label{sec:rep}

As we have said, the physical device used to create the hologram may only modulate either the amplitude or the phase of the light, but not both simultaneously. A DMD consists of a large array of micro-mechanical mirrors in which each mirror can be switched between two different angles, which we refer to as `0' and `1'. A mirror in the `1' position reflects light through the remainder of the optical setup whereas a mirror in the `0' position deflects light towards a beam stop, thus acting as a binary amplitude modulator. A typical full frame rate for such a device is $20\,\mathrm{kHz}$. In contrast, PSLMs are 1-2 orders of magnitude slower, but they offer the advantage of improved control over the hologram, since despite no amplitude control, they permit a quasi-continuous modulation of the phase between $0$ and $2\pi$. In a liquid crystal PSLM, an applied voltage across each of the pixels of the device causes the phase of the light traversing that pixel to be modulated by an amount proportional to that voltage. For a digitally controlled PSLM, the standard response time is $10\,\mathrm{ms}$, and $256$ phase levels are the norm.

For a single trap, we see from equation \ref{eq:hologram} that the amplitude required is constant, and so modulating the phase of the hologram $H(x,y)$ is sufficient to reproduce the trap in the image plane when Fourier transformed. For a PSLM, the approach is thus to simply round the phase value required at each pixel to the nearest value that the device can produce. With over $200$ phase levels, this is effectively just directly displaying the phase value required for each pixel. The theoretical maximum power in this case is $100\%$ of the power incident on the device. 

This is in contrast to the binary amplitude modulating DMD. For the latter, the simplest way to represent the hologram is to map all pixels whose phase is between $-\pi/2$ and $\pi/2$ to the `1' state, and all other pixels to the `0' state. This results in the maximum possible amount of optical power being directed into the trap, since all pixels in the `1' state interfere constructively, and all those that would interfere destructively are in state `0'. The result is a top hat grating where the fraction of power in the $n^{th}$ diffraction order is given by $\frac{1}{4}\mathrm{sinc}^2(n\pi/2)$. The trap is produced in the $+1$ diffraction order, with a theoretical maximum power of $1/\pi^2 \approx 10.1\%$. We necessarily also have an equivalent trap in the $-1$ order, separated by the same distance as our desired trap from a bright $0$th order spot. We see this in the obtained image planes depicted in figure \ref{fig:setup}. Further, there will be small fractions of the power in higher diffraction orders. For either type of modulator, we refer to this simple hologram generation method as the rounding algorithm, though we note that for a PSLM there is no real algorithm involved and the effects of rounding are negligible.

For the case of multiple traps, the problem of representing the complex hologram with an amplitude-only or phase-only modulator is more difficult. Due to the nature of Fourier transforms, the hologram required is effectively the sum of the holograms required for each individual trap. However, the technical limitations of the hologram representation on either a DMD or a PSLM results in additional unwanted ghost traps and drastic variations in intensities of the traps. These artifacts arise from two sources: the quantisation of continuous pixel values to discrete values, and the fact that either any pixel amplitudes outside the range $(0,1)$ are truncated (DMD) or amplitude variations are ignored altogether (PSLM).

\section{Algorithms for Improved Hologram Generation}

To overcome the problems faced in representing holograms for multiple traps, we discuss two more sophisticated algorithms for hologram computation. The first is a dithering algorithm which we refer to as error diffusion dithering, and the second is an iterative Fourier transform algorithm (IFTA) called the mixed-region amplitude freedom (MRAF) algorithm. We now describe each of these algorithms.

\subsection{Floyd-Steinberg Error Diffusion Dithering}

\begin{figure*}[tb] 
	\centering
	\includegraphics[width=0.85\textwidth]{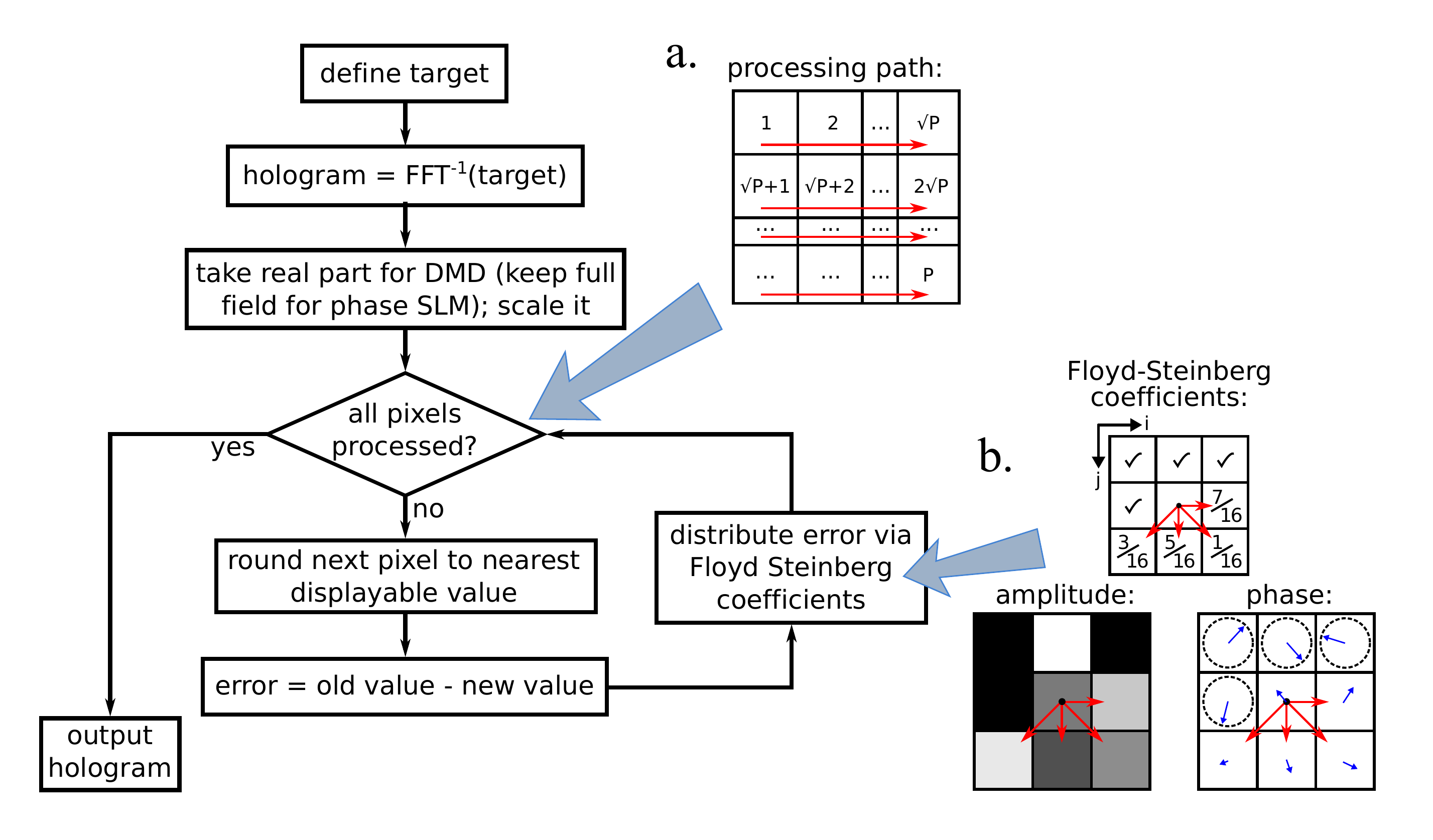}
	\caption{Error diffusion dithering algorithm for hologram generation. (a) The order in which pixels are processed in the algorithm is a sequence of scan lines across the array. (b) The error in rounding a given pixel is shared amongst the neighbouring pixels in the ratios shown. The pixel with a dot is currently being processed. The error distribution is represented pictorially for an amplitude and a phase modulator. In the amplitude case, the greyscale pixel values for the the real part of the hologram field are rounded to zero (black) or one (white), and the scalar error is shared to the neighbours. In the phase case, the vector corresponding to the hologram field at each pixel is approximated by a vector of amplitude one with the nearest phase displayable by the PSLM, and the vector error relative to the \textit{original} vector (before setting the amplitude to one) is shared to the neighbours.}
	\label{fig:errDiff}
\end{figure*}

Dithering algorithms seek to minimise artifacts by mimicking continuous greyscales on a discretised phase-only or amplitude-only modulator. For each pixel, the desired pixel value is set to the closest value that can be displayed by the device. The error, that is the difference to the desired value, is then compensated for using some of the neighbouring pixels. For example, on a DMD, a $50\%$ greyscale could be represented by alternating pixel values between `0' and `1'. 

We use an error diffusion dithering algorithm based on Floyd-Steinberg \cite{floyd}. For either type of modulator, the first step is to create a target image $F$. We set each pixel containing a trap to $\exp(i\theta_n)/\sqrt N$ where $\theta_n$ is a random phase and $N$ is the total number of traps. Next, we perform an inverse Fourier transform on the array to find the hologram $H$. For a DMD, we take the real part of the complex-valued hologram field and scale it to the range $[0,1]$. For a PSLM, we keep the full complex field and scale it to have a maximum intensity of $1$. The pixels are then processed according to a path consisting of a sequence of scan lines, illustrated in figure \ref{fig:errDiff}a. The error is calculated as ``desired value" minus ``displayed value". This error is distributed to the connected neighbours which have not yet been processed, with the error coefficients (fractions of the error sent to each pixel) proposed in \cite{floyd}. These coefficients are shown in figure \ref{fig:errDiff}b. For example, processing pixel $(i,j)$ gives $5/16$ of its error to pixel $(i+1,j)$, which is the next pixel to be processed. When processing this pixel, the input value to be rounded is then equal to its original value \textit{plus} the error fraction given to it from pixel $(i,j)$ (and also pixels $(i,j-1)$, $(i+1,j-1)$ and $(i+2,j-1)$, which would already have been processed).

This process of error distribution is illustrated pictorially in figure \ref{fig:errDiff}b. For an amplitude modulator, the error is a scalar: the difference between the (scaled) real part of the hologram field at that pixel and whichever is closer to this out of the available levels `0' and `1'. For a phase modulator, the vector corresponding to the complex value of the hologram field at that pixel is approximated by the equivalent vector of unit length (i.e. a vector with the same phase, rounded to the nearest available phase level, but now with amplitude $1$). Then the \textit{vector} error in approximating this pixel is distributed. For the next pixel to be processed, the starting point is the original vector for that pixel plus the vectors corresponding to the error fractions accumulated from neighbours that have already been processed.

\subsection{Mixed Region Amplitude Freedom (MRAF)}
 
\begin{figure*}[tb] 
	\centering
	\includegraphics[width=0.85\textwidth]{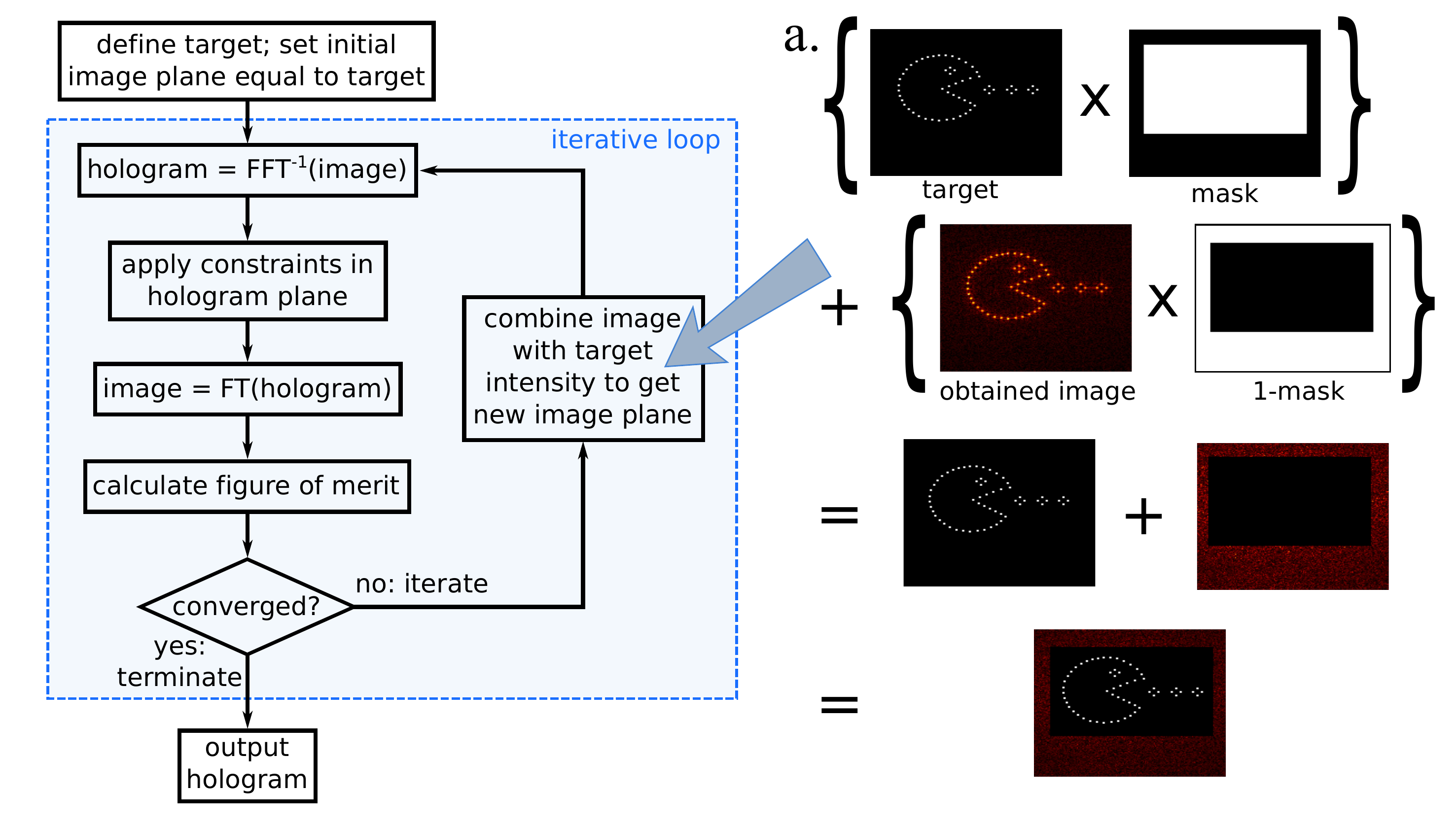}
	\caption{General inverse Fourier Transform algorithm for hologram generation. The iterative loop is repeated until the figure of merit reaches some criteria, signalling that either all or part of the image plane intensity closely enough resembles that of the target image plane. (a) Combining the generated and target intensities in the specific case of the mixed region amplitude freedom (MRAF) algorithm. The signal region is set to the target intensity, while the noise region is left equal to the generated intensity.}
	\label{fig:mraf}
\end{figure*}

The MRAF algorithm is an example of a class of algorithms known as Iterative Fourier Transform Algorithms (IFTAs), the best known of which is the Gerchberg-Saxton algorithm \cite{Optik.35.237}. All IFTAs are broadly similar, in that they exploit phase freedom in the image plane in order to minimise the difference between the desired and obtained intensity distribution in the output image.

The first step in an IFTA is to define a target image, as in the dithering algorithm already discussed. Next comes the iterative part of the algorithm. The image plane is inverse Fourier transformed to get the hologram, and the relevant constraints in the hologram plane are applied. For a DMD this amounts to taking the real part of the hologram and rounding to `0' or `'1', while for a PSLM the amplitude information is discarded and the phase is rounded to the nearest available level. This newly constrained hologram plane is Fourier transformed to get the first iteration of the image. A figure of merit is calculated, comparing the obtained image to the target. If the figure of merit has converged sufficiently, or passed some threshold value, the iterative process is ended. Otherwise, the image is combined in some way with the desired target, and the iterative process is started again. The generic IFTA is illustrated in figure \ref{fig:mraf}.

The step that defines MRAF in particular is the way in which the current iteration's image plane is combined with the target. In the Gerchberg-Saxton algorithm, the intensity everywhere is set to the desired intensity at that location (while the phases are left unchanged so that they may evolve as needed). In contrast, in MRAF we divide the image plane into two regions: one that we consider to be important and refer to as the signal region, and one that we consider unimportant and refer to as the noise region. The goal is to improve the accuracy in the signal region at the expense of less accuracy in the noise region. As such, the algorithm uses a mixing parameter $p$ to control the fraction of power in each of the two subsets of the image plane \cite{Pasienski:08}. Mathematically, for each pixel $(i^\prime,j^\prime)$ in the image plane we define a region masking matrix:

\begin{equation}
	\label{mask}
	\mathbb{M}_{\mathrm{reg}}(i^\prime,j^\prime) = 
	\begin{cases}
      		1, & \text{if}\ (i^\prime,j^\prime)\in\text{signal region}\ \\
      		0, & \text{otherwise.}
    	\end{cases}
\end{equation}
Then the $r$th iteration's image plane $|F^{(r)}|\mathrm{e}^{\mathrm{i} \theta_{\mathrm{im}}^{(r)}}$ is combined with the target intensity distribution $|F_0|^2$ according to

\begin{equation}
\label{eq:mraf}
F^{(r)} = \big{(}\sqrt{p}|F_0|\odot\mathbb{M}_{\mathrm{reg}} + \sqrt{(1-p)} |{F^{(r)}}| \odot(1-\mathbb{M}_{\mathrm{reg}}) \big{)} \mathrm{e}^{\mathrm{i} \theta_{\mathrm{im}}^{(r)}},
 \end{equation}
where $F^{(r)}$ is the calculated image plane after the $r$th iteration, and $\odot$ represents element-wise multiplication. In effect, the amplitudes of the pixels in the signal region are set back to those of the corresponding pixels in the target, while those of the pixels in the noise region are left unchanged, with overall multiplicative factors from the mixing parameter. The phases are left unchanged in both regions.  This step is illustrated visually in figure \ref{fig:mraf}a. By allowing more of the power to be directed into the noise region, i.e. by lowering the parameter $p$, the accuracy achievable in the signal region may be improved.

 \section{Comparison of the Algorithms}
 
All of the algorithms have similar capabilities: they can be used to position a large number of traps in two dimensions (and indeed this can be extended by a small amount into three dimensions by superposing a lens pattern on the calculated hologram). Furthermore, they all suffer from the same set of problems as a result of the physical limitations of the modulator: variation of the optical power between traps, loss of power in the form of noise and ghost traps, and when using a DMD, further loss of power into unwanted diffraction orders. We evaluated the performance of each of the algorithms with a numerical simulation for a periodic square lattice of traps, with $4$ dark pixels between each trap pixel. This type of grid layout is an arrangement of particular interest for many applications of optical tweezers, including quantum computing. 
 
\begin{figure*}
	\centering
	\includegraphics[width=0.9\textwidth]{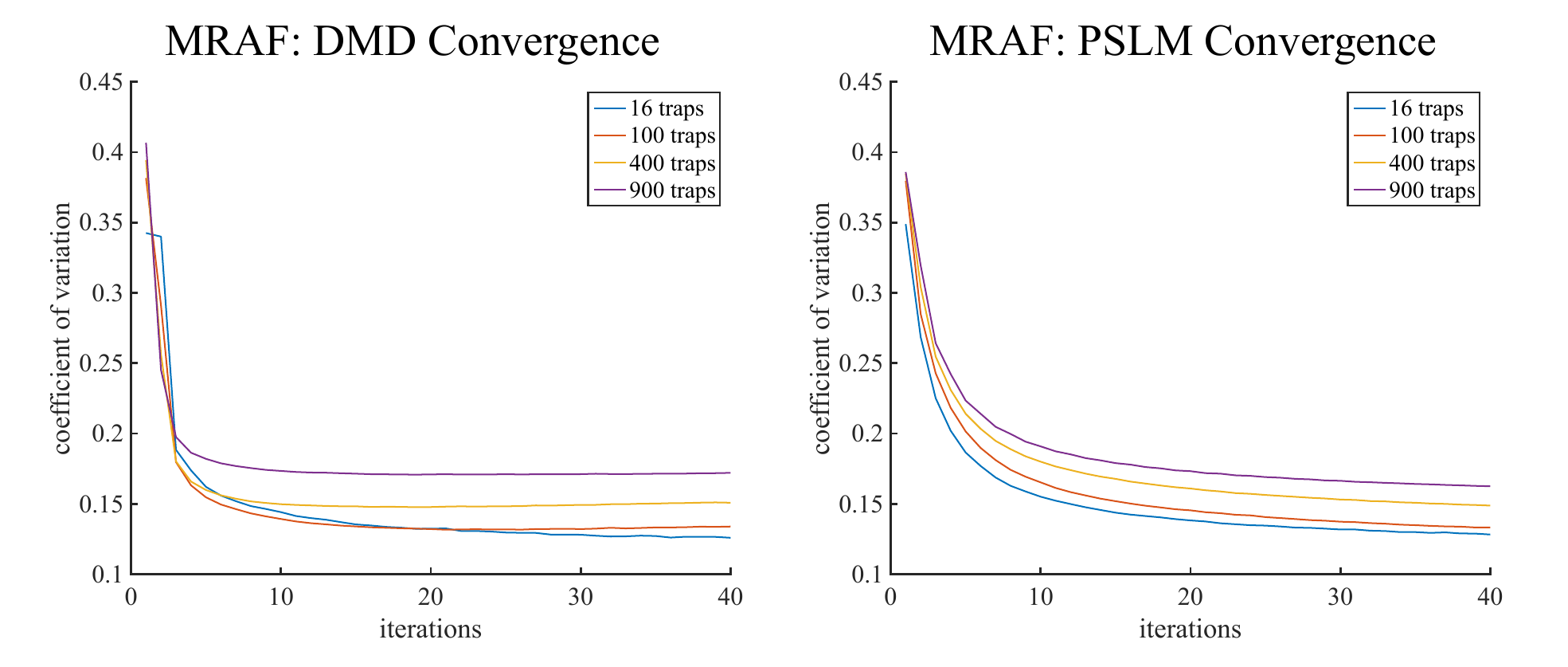}
	\caption{Error in approximating the desired image plane as a function of number of iterations of the MRAF algorithm, calculated via equation \ref{eq:acc} for a binary DMD (left) and a PSLM with 256 levels (right). In each case the traps were arranged in a square grid. The PSLM converges more consistently, but the overall degree of convergence is similar in each case. The rate of convergence is approximately independent of the number of traps (particularly for the PSLM); after approximately 5 iterations we get diminishing returns on further iterations, and by 20 iterations the further improvements are negligible. The results here were averaged over 1250 runs of the simulation for each number of traps.}
	\label{fig:mrafConvergence}
\end{figure*} 
 
The relevant things to compare are: efficiency (what fraction of the power ends up in the traps), speed of calculation, and accuracy (how much power variation there is between traps). Further, we consider how the efficiency and accuracy scale with the number of traps. In the case of MRAF, rather than using a figure of merit to determine when iterations should be ceased, we instead artificially stop the iterations after some number and use the figure of merit to investigate how the image accuracy is affected by the number of iterations we perform. 

For MRAF, we use a signal area consisting of $200\times200$ pixels, or approximately $15\%$ of the total area of $512\times512$ pixels. We use a mixing parameter of $p=0.7$ for our comparison. The authors of \cite{mraf2} find that maximum accuracy may, in general, be achieved for a parameter of $0.4$, but we find for our trap grids that up to $0.7$ makes little difference to the accuracy whilst almost doubling the efficiency.

We begin by considering computation speed. This is hard to quantify, since it depends on the quality of code and the machine on which it is executed. Thus we quote speeds as the algorithmic complexity, assuming that the only significant time costs are for Fourier transforms, Floyd-Steinberg error distribution, and the element-wise matrix multiplication used in the masking of MRAF. For a 2D array of $P$ pixels in total, the time for a two dimensional discrete Fourier transform scales as $P\log P$, while error distribution and element-wise matrix multiplication both scale as $P$. The relative speeds of the algorithm are shown in table \ref{table:comparison}. Because the MRAF algorithm is iterative with Fourier transforms every iteration, and the Fourier transform scales least favourably with number of pixels, MRAF is by far the slowest algorithm. 

\begin{figure*}
	\centering
	\includegraphics[width = 0.9\textwidth]{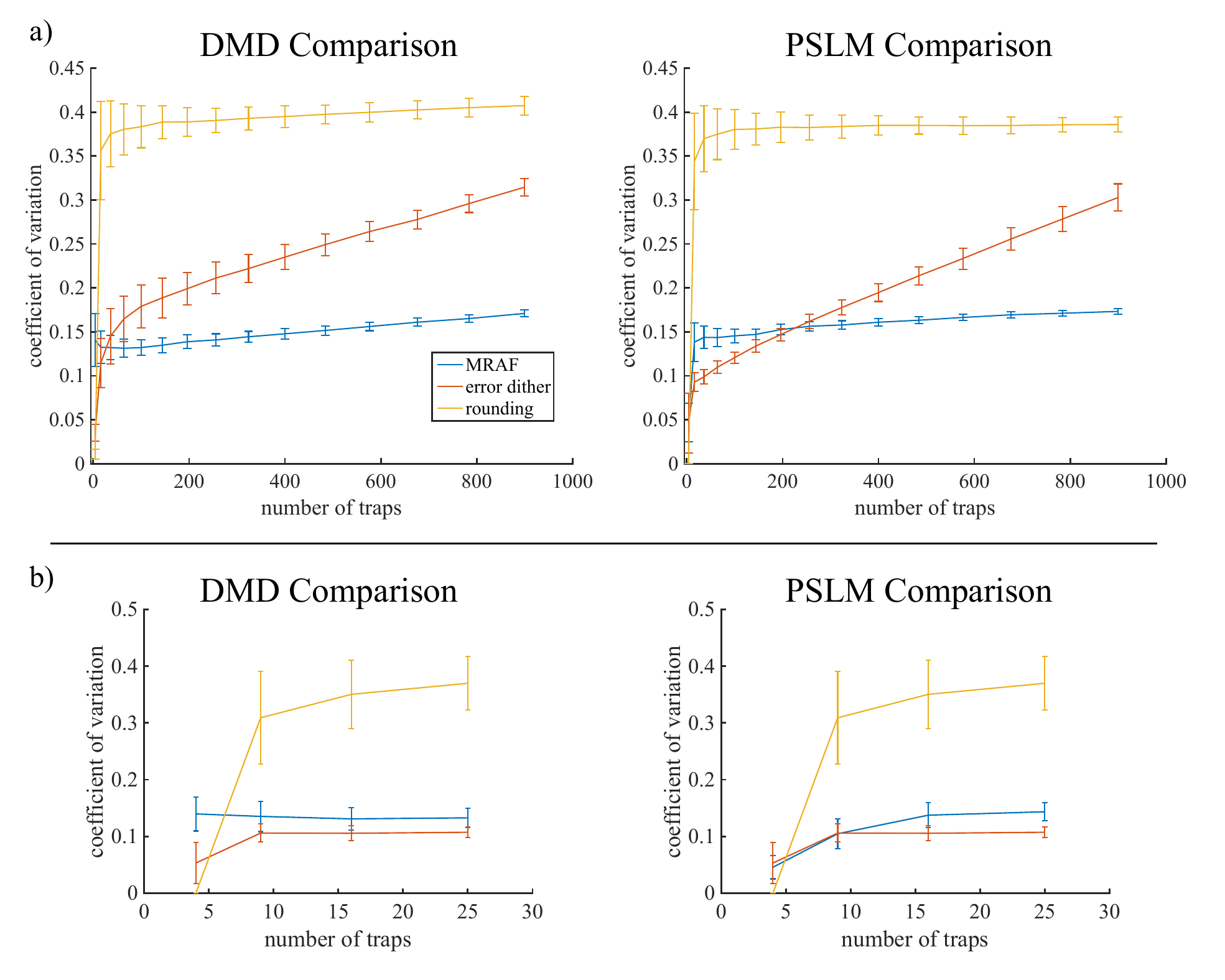}
	\caption{Accuracy of each of the algorithms as a function of number of traps. Each data point is the median average over 1250 runs of the simulation, with error bars at the inter-quartile range. a) $2 \times 2$ grid up to $30 \times 30$ grid of traps. We see that for large numbers of traps, error diffusion dithering continues to get worse as more traps are added, while above 25 traps, MRAF and rounding barely change. Regardless of the type of modulator, MRAF performs best for large numbers of traps, though the crossover between dithering and MRAF occurs at a much smaller trap number for an amplitude modulator. b) A closer look at grids from $2 \times 2$ up to $5 \times 5$. For a $2 \times 2$ grid the rounding algorithm outperforms the others, and all algorithms do substantially better.}
	\label{fig:accNumTraps}
\end{figure*}

Knowing that MRAF is comparatively slow as a result of its iterative nature, we now ask just how many iterations are necessary. As a measure of accuracy, we use the coefficient of variation $c_{\mathrm{v}}$ of the intensities at the trap locations:

\begin{equation}
\label{eq:acc}
c_{\mathrm{v}} = \frac{\sigma}{\langle I \rangle} = \frac{\sqrt{ \frac{1}{N} \sum\limits_{n \in \mathrm{traps}} (I_n - \langle I \rangle)^2 }}{\langle I \rangle},
\end{equation}

i.e. the standard deviation of the intensities divided by the mean trap intensity. Here $N$ is the number of traps in the grid, $I_n$ is the intensity of the $n$th trap, and $\langle I\rangle$ is the mean trap intensity. Empirically we find that all the algorithms reproduce the dark pixels between traps well, and struggle with producing equal powers between the traps. Sometimes unwanted ghost traps are produced, but these are far enough away from the target traps that they don't cause any problems besides loss of power into the target traps. Hence considering only the trap pixels in our metric makes sense. Further, this metric has the advantage of being an intrinsic property, and one which is completely independent of the efficiency of an algorithm and the total input light power.

Figure \ref{fig:mrafConvergence} shows how the coefficient of variation varies with numbers of iterations of the MRAF algorithm. We see that after 5 iterations, the speed of convergence starts to drop rapidly and we get diminishing returns. By around 15 to 20 iterations, further improvements are fairly negligible. 

\begin{figure*}
	\centering
	\includegraphics[width=0.8\textwidth]{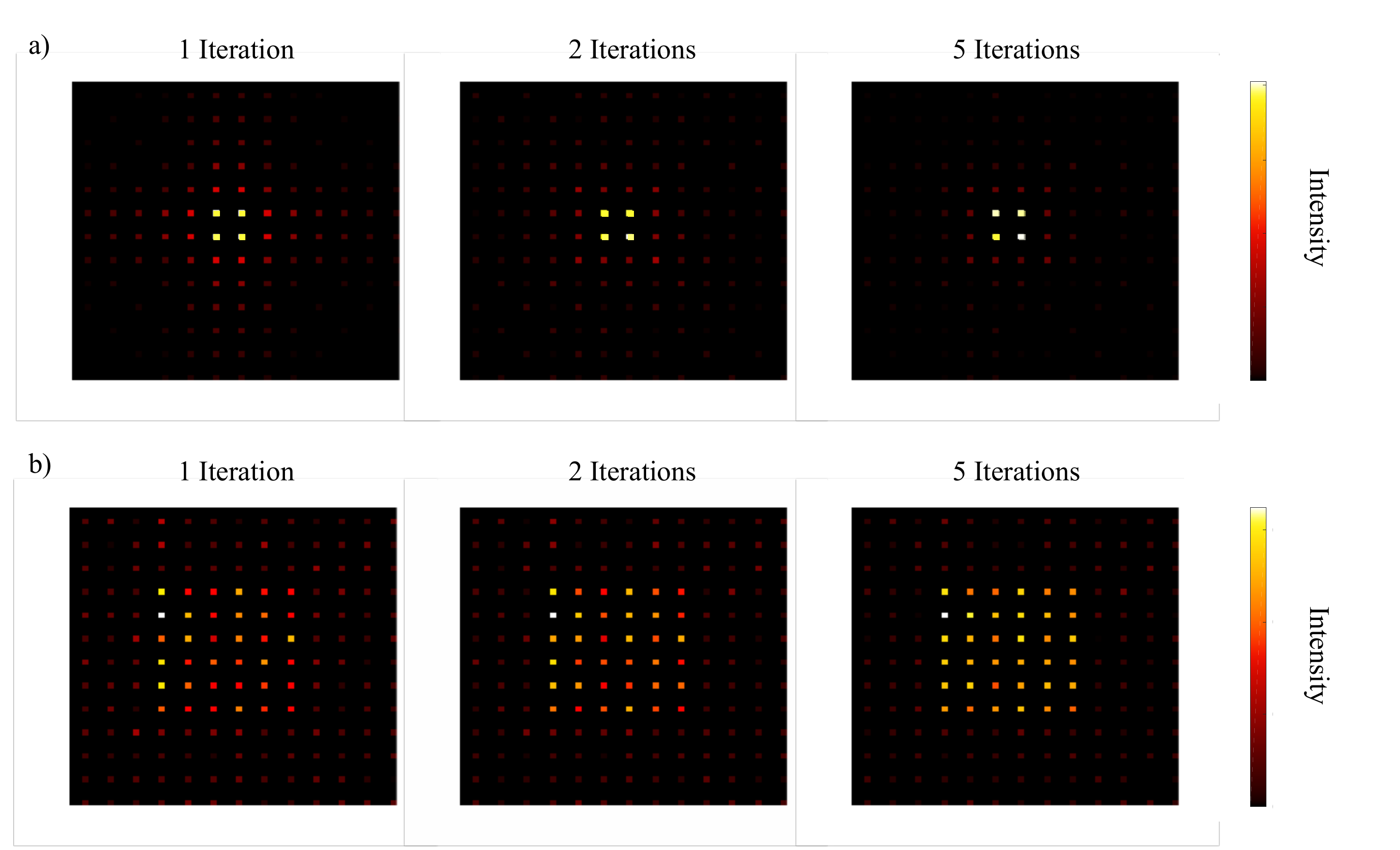}
	\caption{Convergence of the MRAF algorithm for a) a 2$\times2$ grid of traps, b) a $6\times6$ grid of traps. Only the signal region, and not the noise region, is shown here.}
	\label{fig:mrafIts}
\end{figure*}

We now consider the accuracy of the MRAF algorithm, as compared to each of error diffusion dithering and simple rounding. For this comparison, we use $20$ iterations of MRAF. The accuracy as calculated by equation \ref{eq:acc} for each algorithm, as a function of number of traps, is shown in figure \ref{fig:accNumTraps}. Overall, we see that the dithering algorithm performs best for small numbers of traps, but above a certain point its performance is overtaken by that of the MRAF algorithm. For a DMD, this point is for a $6\times6$ grid or greater, while for a PSLM MRAF only performs better for $16\times16$ grids or bigger.

\begin{table*}
\begin{center}
\begin{tabular}{ |c|c|c|c|c|} 
\hline
& & rounding & dithering & MRAF \\
\hline
& computational speed & $P\mathrm{log}P$ & $P\mathrm{log}P + P$ & $\mathrm{iterations}\times(P\mathrm{log}P + P)$ \\ 
\hline
& efficiency & 0.9 & 0.4 & 0.7 \\ 
\cline{2-5}
PSLM & accuracy ($4\times4$ traps) & 0.35 & 0.10 & 0.14 \\ 
\cline{2-5}
& accuracy ($20\times20$ traps) & 0.38 & 0.19 & 0.16 \\ 
\hline
& efficiency & 0.1 & 0.04 & 0.07 \\ 
\cline{2-5}
DMD & accuracy ($4\times4$ traps) & 0.35 & 0.11 & 0.13 \\ 
\cline{2-5}
& accuracy ($20\times20$ traps) & 0.39 & 0.23 & 0.15 \\ 
\hline
\end{tabular}
\caption{A comparison of the speed and efficiency of each of the three algorithms. $P$ is the number of pixels used in the calculation. MRAF is substantially slower than the other two algorithms. Dithering is the least efficient. For all three algorithms, we get a 90\% loss in power for a DMD as compared to a PSLM. Note that the efficiency for MRAF is imposed by our choice of $p=0.7$, as this maximises the accuracy.}
\label{table:comparison}
\end{center}
\end{table*}

At very small trap numbers, we observe some interesting behaviour. All of the algorithms do substantially better below a $4 \times 4$ grid, and further, MRAF and error diffusion dithering are outperformed by the simple rounding algorithm. The reason for this becomes clear when we consider how we are calculating our errors, and look at the signal region of the image plane generated by MRAF, say, for the first few iterative steps. This is shown for a $2 \times 2$ grid in figure \ref{fig:mrafIts}a. For the first iteration, the traps look good, but there are many additional ghost traps within the signal region. Overall, the region gets improved with subsequent iterations. However, the variation between trap intensities gets worse, because the algorithm does not know that the trap pixels are any more important to us than the non trap pixels. As a result, the trap pixels are sacrificed to some extent to enable the overall similarity of the signal region to the target signal region to improve with more iterations, but the coefficient of variation for the traps increases. This behaviour is not seen for larger trap numbers, as then the initial iteration does not do a good job of creating equal intensity traps, and so the biggest improvement the algorithm can make to the signal region is to try and equalise these intensities. This is shown for a $6 \times 6$ grid of traps in figure \ref{fig:mrafIts}b.

Efficiency of the algorithms, i.e. how much of the input light ends up in the traps, is also a practical concern. For most applications, input power will be restricted in some way, and there may be problems associated with power being directed to unwanted locations. The efficiencies of each of the algorithms are summarised in table \ref{table:comparison}. In each case, the efficiencies for a DMD are an order of magnitude worse than those for a PSLM, as explained in section \ref{sec:rep}. The dithering algorithm is less efficient than the others for both types of modulator. The reason for this is that, in essence, the dithering process pushes the noise into high frequency components, which are directed away from the traps. The MRAF efficiency is not particularly good for comparison, as this is artificially imposed by our choice of a mixing parameter $p=0.7$. We note that by increasing the value of $p$, the efficiency of the algorithm may be increased at the expense of the accuracy.

\section{Conclusion}

In conclusion, we have shown that either a DMD or a PSLM may be used to holographically generate arrays of light spots ideally suited for optical tweezers. The DMD is fast, but inefficient in comparison to the slower PSLM. We have demonstrated a number of algorithms for hologram computation, and discussed their merits. For large numbers of traps ($6\times6$ grid or greater for a DMD, or $16\times16$ grid or greater for a PSLM), MRAF gives the most accurate results. For smaller numbers of traps, error diffusion dithering performs better. The trade-off is an almost $50\%$ loss of power compared to MRAF, and the lower consistency of the accuracy, particularly when used with an amplitude modulator. For very small numbers of traps ($2\times2$ grid) the simple rounding algorithm works better than either of the more sophisticated algorithms.

\end{document}